\documentclass[acmsmall]{acmart}

\usepackage{enumitem}
\usepackage{subcaption}
\usepackage{makecell} 

\AtBeginDocument{%
  }

\setcopyright{cc}
\setcctype{by}
\acmJournal{PACMHCI}
\acmYear{2026} \acmVolume{10} \acmNumber{3} \acmArticle{ETRA003}
\acmMonth{5} \acmDOI{10.1145/3806017}

\begin{document}

\title[A Scoping Review and Case Study of Crowdsourced Webcam Eye Tracking in AI Interviews]{What Shapes Participant Data Quality? A Scoping Review and Case Study of Crowdsourced Webcam Eye Tracking in AI Interviews}

\author{Ka Hei Carrie Lau}
\email{carrie.lau@tum.de}
\orcid{0009-0005-8838-3230}
\affiliation{%
  \institution{Chair of Human-Centered Technologies for Learning, Munich Center for Machine Learning (MCML), Technical University of Munich}
  \city{Munich}
  \country{Germany}
}

\author{Enkelejda Kasneci}
\email{enkelejda.kasneci@tum.de}
\orcid{0000-0003-3146-4484}
\affiliation{%
\institution{Chair of Human-Centered
Technologies for Learning, Technical University of Munich}
\city{Munich}
\country{Germany}}

\begin{abstract}

Webcam-based eye tracking is a cost-effective, scalable method for remote research that effectively reaches broader populations. However, uncontrolled environments and hardware diversity lead to inconsistent data quality in crowdsourcing. To assess current practices, we conducted a scoping review of crowdsourced eye-tracking from 2011--2025. The review confirms fragmented reporting and a lack of established quality benchmarks. To address this lack of predictive insight, we conducted a case study on AI fairness interviews ($N=205$) using the RealEye platform. Applying Ordered Logistic Regression (OLR) to the platform’s quality metric, we found that behavioral and technical factors significantly predict data quality. Specifically, within the RealEye platform, higher fixation counts, shorter sessions, and operating system choice yield significantly higher quality grades. Based on this review and platform-specific predictive insights, we provide actionable recommendations to enhance the reliability, transparency, and replicability of future crowdsourced webcam eye tracking in HCI and behavioral science.

\end{abstract}



\begin{teaserfigure}
  \centering
  \includegraphics[width=\textwidth]{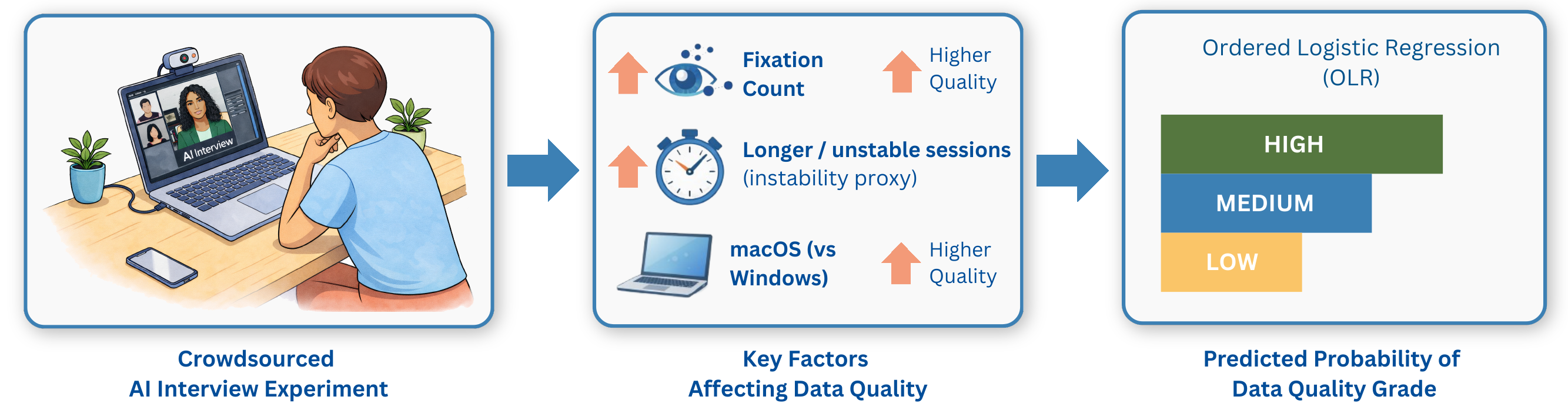}
  \caption{Overview of our approach to evaluating participant data quality in crowdsourced webcam-based eye tracking during AI interviews. Behavioral and device-related factors predict data quality, and an ordered logistic regression (OLR) model is used to estimate the probability of different quality grades. Visuals were generated using GPT-5.3 and refined by the authors.}
  \Description{Overview of our approach to evaluating participant data quality in crowdsourced webcam-based eye tracking during AI interviews. Behavioral and device-related factors predict data quality, and an ordered logistic regression (OLR) model is used to estimate the probability of different quality grades. Visuals were generated using GPT-5.3 and refined by the authors.}
  \label{fig:teaser}
\end{teaserfigure}

\maketitle

\section{Introduction}

Webcam-based eye tracking is a low-cost and scalable method that has democratized eye-tracking and behavioral research~\cite{yang2021webcam, vos2022comparing, patterson2025methodological, prystauka2024online, van2024validation, james2025paradigms}.
By leveraging participants’ consumer webcams, it enables remote studies and access to diverse populations~\cite{bertrand2023dynamics, huang2018quick}, extending the reach of research beyond the laboratory.
Despite advances in webcam eye trackers, researchers still lack a systematic understanding of the reliability of crowdsourced webcam gaze data.
Recent studies have reported inconsistent sampling precision and calibration stability across different hardware configurations~\cite{patterson2025methodological, Heck2023}.

Unlike infrared (IR) eye trackers, webcam-based methods operate under uncontrolled conditions~\cite{banki2022comparing, semmelmann2018online}.
Variations in lighting~\cite{yang2021webcam}, camera quality~\cite{kim2017bubbleview}, hardware performance~\cite{thilderkvist2022current}, and participant behavior~\cite{kandel2025assessing} introduce noise that reduces spatial and temporal precision~\cite{juantorena2023web, gagne2023run, thilderkvist2022current}.
Gaze estimates can lag by about 300~milliseconds (ms) relative to IR benchmarks~\cite{slim2023moving} and vary in sampling frequency or positional accuracy across devices~\cite{ribeiro2023webqamgaze}.

However, prior research has rarely quantified how participant behavior and device configuration affect data reliability.
Individual studies have evaluated specific factors such as head movement~\cite{sharafi2020practical, thilderkvist2022current} and screen distance~\cite{juantorena2023web}, but few have statistically modeled these effects, particularly for commercial platforms like RealEye\footnote{\url{https://www.realeye.io/}, last accessed 11~February~2026}.
Without such a predictive framework, researchers cannot proactively identify or mitigate reliability issues in remote eye-tracking data, thereby limiting progress toward robust, replicable findings. Although our empirical analysis uses data provided by the RealEye platform, the factors we examine, such as fixation count, test duration, and operating system, are not specific to RealEye. Rather, they reflect general challenges of unsupervised, crowdsourced, webcam-based eye tracking, and our findings can help researchers understand the factors that influence data quality in such settings. To structure this research, we investigate the following research questions (RQs):

\begin{quote}
\textbf{RQ1.} What methodological and validation practices have defined the development of webcam-based and crowdsourced eye-tracking research?

\textbf{RQ2.} Which behavioral and technical factors predict data quality in crowdsourced webcam-based eye-tracking?
\end{quote}

We address these questions through a scoping review of webcam-based and crowdsourced eye-tracking research from 2011 to 2025 and an empirical analysis of participant-level data quality.
The review identifies three research areas: system development, validation, and application. Additionally, it highlights research gaps in quality reporting and predictive modeling.
Our empirical analysis uses data from 205 participants in an artificial intelligence (AI) fairness interview study on the RealEye platform. Using ordered logistic regression (OLR), we investigated how behavioral and technical factors predicted the quality of gaze data. We chose this setting because AI interviews are a socially interactive and attention-demanding task that mirrors realistic webcam-based interactions. This makes them a suitable context for evaluating data quality under unsupervised conditions. To our knowledge, this study is the first to combine a scoping review with an empirical, crowdsourced eye-tracking analysis in the emerging field of AI fairness.

Our findings show that webcam gaze data quality varies systematically rather than randomly. A higher total number of detected fixations per participant session and shorter test durations are associated with better data quality, and device-related factors also contribute significantly. These insights help refine guidelines for study design, participant screening, and data quality assessment, advancing methodological understanding of webcam-based eye tracking. To this end, we make two contributions:

\begin{description}
\item[Survey.] We conduct a scoping review of crowdsourced webcam-based eye-tracking research across tasks, platforms, and validation methods. The review summarizes reported accuracy and data-loss measures, outlines current approaches, and identifies gaps in methodological transparency.

\item[Empirical.] We present empirical evidence on factors influencing data quality in crowdsourced webcam-based eye-tracking. Using OLR on a crowdsourced social perception dataset (\textit{N} = 205), we evaluate the relationship between behavioral and technical variables and the likelihood of obtaining high-quality gaze data.
\end{description}

\section{Related Work}

Eye tracking has been applied across extended reality (XR), mobile, and webcam-based platforms, each of which poses distinct methodological challenges. Recent reviews highlight advances in gaze estimation, interaction design, and data analysis driven by low-cost sensors and deep learning~\cite{lei2023end, bozkir2025eyetrackedVR, Plopski2022, adhanom2023, Katsini2020}. However, they also note persistent issues with calibration reliability, data noise, and inconsistent reporting in unconstrained settings.

\paragraph{Methodological standards.}
Prior reviews have proposed guidelines for designing and reporting eye-tracking studies, emphasizing consistent calibration, transparent exclusion criteria, and standardized quality metrics~\cite{CARTER202049, Blascheck2014StateoftheArtOV}. Building on this,~\citet{patterson2025methodological} examined webcam-based and crowdsourced studies, identifying fragmented reporting and recommending structured documentation of sampling rates, calibration accuracy, and exclusion thresholds. Similar issues appear in XR and mobile research, where studies report trade-offs between scalability and precision and recurring concerns about privacy and robustness~\cite{lei2023end, bozkir2025eyetrackedVR}. Altogether, these works underscore the need for standardized reporting frameworks to improve transparency and facilitate future cross-platform comparability.

\paragraph{System comparisons.}
Empirical comparisons between webcam-based and IR eye trackers reveal well-known issues, including spatial inaccuracy and calibration instability~\cite{shehu2021remote}. Although deep learning has improved estimation robustness, webcam systems still face limitations in temporal precision and participant non-compliance~\cite{vos2022comparing, patterson2025methodological}. Most reviews remain descriptive, focusing on hardware or algorithmic performance rather than quantifying how behavioral or contextual factors influence data reliability.

\paragraph{Research gap.}
Despite rapid methodological progress, validation and reporting practices have not kept pace. Existing reviews focus mainly on system accuracy and hardware performance, but rarely examine how behavioral and technical factors jointly shape webcam-based gaze data quality. As a result, reproducibility remains limited, and cross-platform benchmarks are still missing. To begin addressing this gap, we combine a scoping review with an empirical modeling analysis. The review summarizes current methods, validation strategies, and applications in webcam-based and crowdsourced eye tracking. The modeling analysis uses an existing RealEye dataset to explore how participant behavior and device factors relate to data quality, offering a first step toward more general, platform-independent models.

\section{Scoping Review}

\begin{figure}[!ht]
\centering
\begin{subfigure}{0.52\textwidth}
\centering
\includegraphics[width=\textwidth, keepaspectratio]{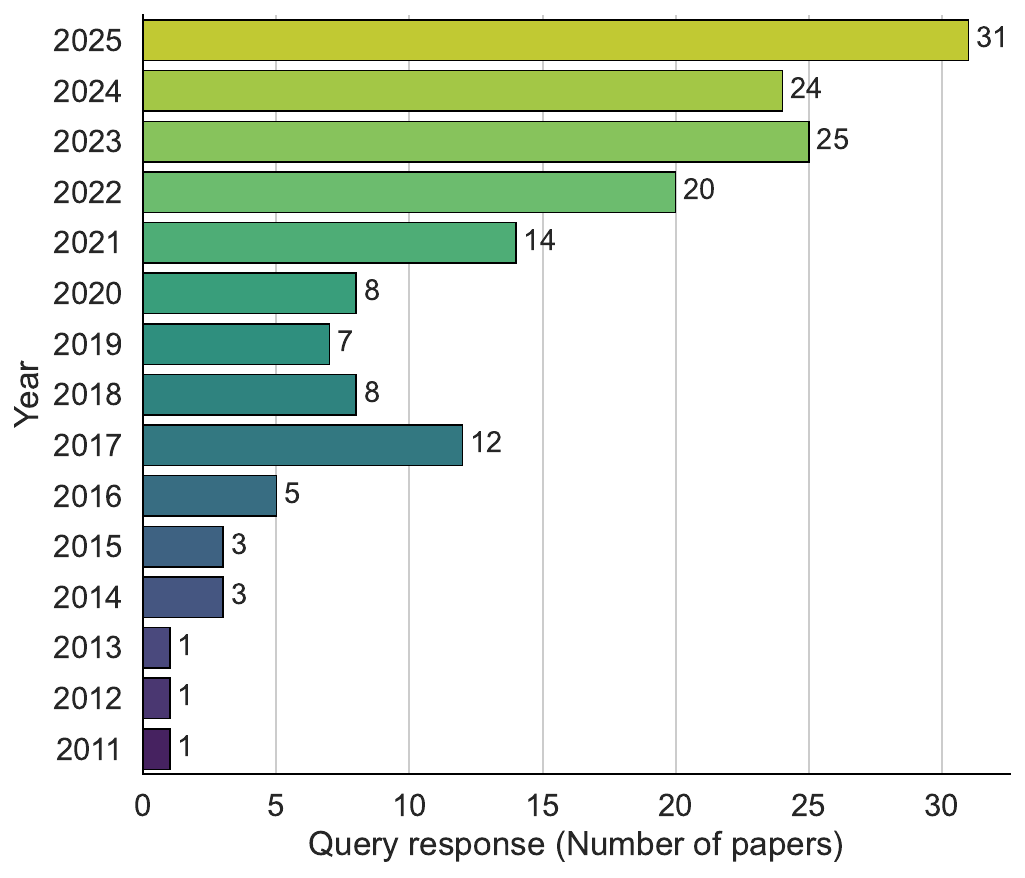}
\caption{Publications returned by the scoping review query (2011–2025).}
\label{fig:main_query_distribution}
\end{subfigure}
\hfill
\begin{subfigure}{0.46\textwidth}
\centering
\includegraphics[width=\textwidth, keepaspectratio]{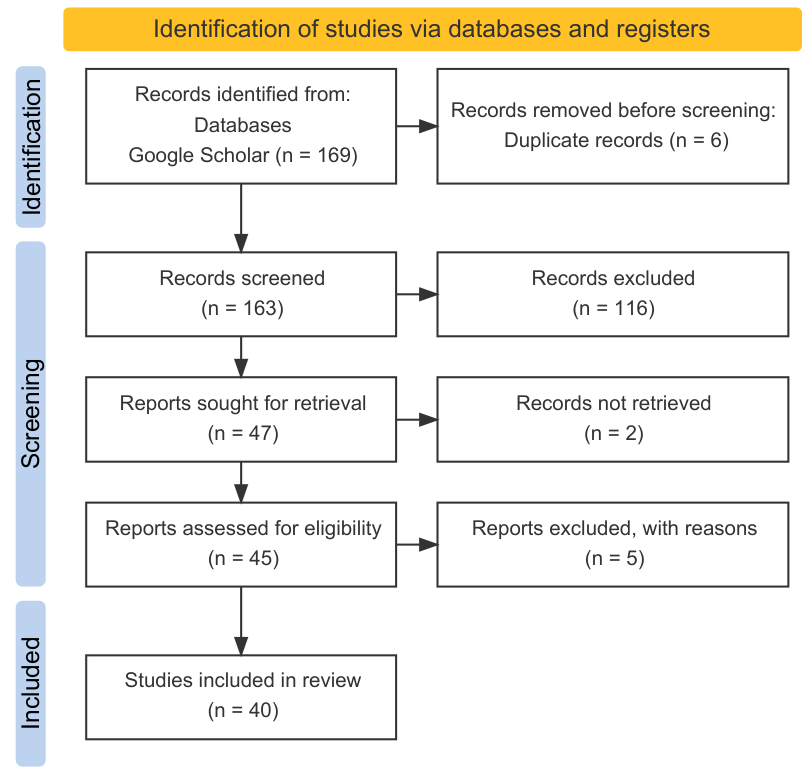}
\caption{PRISMA flow diagram of study identification and screening.}
\label{fig:PRISMAflow}
\end{subfigure}
\caption{Overview of the scoping review process. (a) Number of publications returned by the query. After merging and removing duplicates, 163 unique records were screened and 40 studies were included. (b) PRISMA flow diagram of study identification, screening, and inclusion.}
\Description{The figure consists of two panels. The left panel shows a horizontal bar chart of the number of publications returned by a scoping review query for each year from 2011 to 2025. The counts vary by year, with lower numbers in earlier years and higher counts in more recent years, peaking at 31 publications in 2025. The right panel shows a PRISMA flow diagram of the study selection process: 169 records were identified, 6 duplicates removed, 163 records screened, 116 excluded, 47 reports sought for retrieval, 2 not retrieved, 45 assessed for eligibility, 5 excluded, and 40 studies included in the final review.}
\label{fig:scoping_overview}
\end{figure}

To address~\textbf{RQ1}, we searched Google Scholar for publications from 2011 to 2025. We used the query
``crowdsourcing'' AND (``eye tracking'' OR ``eye movement'' OR ``gaze'') AND ``webcam'' AND (``data quality'').
We did not restrict venues to avoid bias toward a single community, as webcam-based eye tracking spans human-computer interaction (HCI), psychology, computer vision, and marketing. We began with five survey papers as seed articles~\cite{bozkir2025eyetrackedVR,Katsini2020,patterson2025methodological,Plopski2022,shehu2021remote} and extended the set with backward and forward citation chasing. The database search identified 169 records. After removing duplicates (n=6), 163 unique records were screened as presented in Figure~\ref{fig:main_query_distribution}, and 40 papers met the inclusion criteria. Figure~\ref{fig:PRISMAflow} summarizes the identification, screening, and inclusion process. We included peer-reviewed, English-language studies on webcam-based or crowdsourced eye tracking that reported empirical findings or methodological evaluations. We excluded theses, editorials, position papers, tool-only notes, and non-English articles. Two researchers independently screened titles and abstracts and assessed full texts when necessary. Inter-rater reliability for screening decisions was substantial (Cohen’s $\kappa = 0.78$). Disagreements were resolved through discussion, resulting in a final set of 40 included studies.

Our scoping review led to the identification of three main research directions that trace the development of webcam-based and crowdsourced eye-tracking research: (1)~\textbf{Methodology}, which focuses on developing and improving the technology; (2)~\textbf{Validity}, which establishes empirical trustworthiness through comparison with laboratory-based systems; and (3)~\textbf{Application}, which applies webcam- and crowdsourced eye-tracking approaches to investigate behavioral, cognitive, and other research questions across different application domains.

\subsection{Methodology}
This line of research develops the core infrastructure for low-cost and webcam-based eye tracking. Across the literature, three major directions emerge: (1)~\textbf{system and algorithm development}, (2)~\textbf{dataset contribution}, and (3)~\textbf{calibration and pipeline improvement}. These advancements reduce the costs and expertise required for traditional eye-tracking setups, enabling scalable studies to be conducted outside laboratory environments.

\paragraph{System and algorithm development.}

Methodological work primarily focuses on systems and algorithms that enable scalable webcam-based gaze tracking. There are two main approaches:~\textbf{direct webcam eye tracking} and \textbf{attention proxy methods}. Early work that uses \textit{direct webcam eye tracking} includes TurkerGaze~\cite{xu2015turkergaze}, which demonstrated crowdsourced gaze collection via Amazon Mechanical Turk, while WebGazer.js~\cite{papoutsaki2016} introduced an open-source, browser-based library for real-time gaze estimation using implicit calibration. WebGazer has since been integrated into experimental frameworks like jsPsych~\cite{vos2022comparing, yang2021webcam, james2025paradigms, ribeiro2023webqamgaze, juantorena2023web} and Gorilla~\cite{prystauka2024online, bogdan2024investigating}. SearchGazer~\cite{papoutsaki2017searchgazer} extended this approach for search tasks, improving drift correction through implicit interactions. Other webcam systems include UnitEye~\cite{wagner2024uniteye} for 3D environments and Raspberry Pi–based deep-learning gaze predictors~\cite{panja2024prediction}. In contrast, \textit{Attention Proxy Methods} such as BubbleView~\cite{kim2017bubbleview}, FocalVid~\cite{shaghaghi2025focalvid}, and TurkEyes~\cite{newman2020turkeyes} approximate visual attention via mouse clicks or paths, providing scalable alternatives when webcam tracking is unavailable or privacy-sensitive.

\paragraph{Dataset contribution.}
A second key direction involves building datasets for training, validation, and benchmarking of remote eye-tracking and attention models. Such datasets support deep-learning development and enable performance evaluation across diverse hardware and participants. WebQAmGaze~\cite{ribeiro2023webqamgaze} provides a multilingual webcam reading dataset collected with WebGazer and validated against EyeLink lab data. CrowdEyes~\cite{othman2017crowdeyes} uses a low-cost head-mounted webcam and crowdsourcing via CrowdFlower to gather large-scale pupil-localization and fixation-tagging data, extending algorithm training beyond laboratory settings. BubbleView~\cite{kim2017bubbleview} and TurkEyes~\cite{newman2020turkeyes} similarly contribute crowdsourced attention maps and annotation frameworks, serving as both datasets and experimentation platforms. Altogether, these efforts establish the empirical foundation for testing and refining webcam-based gaze estimation methods.

\paragraph{Calibration and pipeline improvement.}
One crucial step toward robust webcam eye tracking is developing calibration methods that are both reliable and user-friendly. Recent research focuses on improving efficiency and personalization to mitigate the noise and variability inherent in real-world webcam data. For example, fast-PACE~\cite{huang2018quick} builds on the Personalized Auto-Calibrating Eye-tracking (PACE) framework, which adapts gaze estimation automatically from natural user interactions such as clicks or typing, thereby reducing the need for explicit calibration.~\citet{saxena2022towards} evaluate streamlined calibration tasks, including brief pursuit routines and device-distance estimation, showing that shorter procedures can maintain accuracy while minimizing participant effort. Overall, these studies demonstrate that adaptive and lightweight calibration strategies are essential for making webcam-based eye tracking both accurate and practical beyond laboratory settings.

\subsection{Validity}
This line of research evaluates the empirical validity of webcam-based eye tracking by comparing its performance with laboratory-grade systems. Three main validation levels emerge: (1)~\textbf{system-level}, assessing technical and measurement equivalence; (2)~\textbf{task-level}, testing behavioral and cognitive replicability; and (3)~\textbf{procedural-level}, defining best practices for reliable data collection under uncontrolled conditions.

\paragraph{System-level validation.}
System-level studies compare webcam-based eye trackers with laboratory-grade IR systems such as EyeLink and Tobii to assess technical performance~\cite{kaduk2024webcam, vos2022comparing, asghari2022can, slim2023moving, patterson2025methodological, hammond2023crowdsourced}. 
Three main limitations were identified: spatial inaccuracy, limited temporal precision, and systematic bias. 
Webcam eye trackers typically show spatial errors of about $3^\circ$ to $4.5^\circ$~\cite{kaduk2024webcam, vos2022comparing, asghari2022can, patterson2025methodological}, and sampling rates of 12–30~Hz restrict the detection of rapid eye movements compared with 100–1000~Hz in IR systems. 
They also exhibit centering bias, for example, gaze clustering near the screen center, and vertical compression, for instance, underestimation along the y-axis due to head motion, lighting variation, and geometric distortion~\cite{kaduk2024webcam, vos2022comparing, slim2023moving}. 
Although convolutional neural network (CNN)–based models can reduce errors to about $2.6^\circ$~\cite{asghari2022can}, they cannot fully overcome the hardware and environmental constraints of consumer webcams. 
Nonetheless, strong correlations between webcam- and lab-based gaze trajectories, around $r = .8$ to $.9$~\cite{kaduk2024webcam, vos2022comparing, slim2023moving}, support webcam-based tracking for attentional and area-of-interest (AOIs) analyses. 
Overall, these findings define the technical boundaries within which webcam eye tracking produces valid behavioral data and motivate our analysis of factors predicting data quality in crowdsourced settings.

\paragraph{Task-level validation.}
This type of validation evaluates whether webcam-based eye tracking can reproduce well-known behavioral and cognitive effects. Across visual attention, language comprehension, and early cognitive development, webcam tracking replicates established gaze patterns, including predictive attention, novelty responses, and real-time language processing, though with smaller effect sizes (40–60\% of laboratory results) and temporal delays of 200–700~ms~\citep{banki2022comparing, bogdan2024investigating, prystauka2024online, swanson2024syntactic, van2024validation, vos2022comparing, slim2023moving}.~\citet{van2024validation} reproduced three classic gaze effects, and \citet{vos2022comparing} replicated verb-aspect processing with only a 50~ms delay. Emotion–attention~\citep{bogdan2024investigating} and infant looking-time~\citep{banki2022comparing} studies confirm broader applicability despite lower spatial precision. Overall, webcam tracking yields valid behavioral measures when analyses target larger AOIs or sustained fixations.

\paragraph{Procedural-level validation.}
This level of research explores how procedural factors, including calibration routines, participant guidance, and recruitment quality, impact data reliability in uncontrolled online environments. 
\citet{patterson2025methodological} highlighted that transparent reporting of calibration thresholds, sampling-rate criteria, and participant instructions is essential for replicable webcam-based eye tracking. 
Likewise, \citet{uittenhove2022lab} compared online and laboratory data collection and found that most data loss arises from participant non-compliance and sample quality rather than the testing environment itself. 
They reported that remote testing introduces only a small decrease in quality and recommended oversampling by about 20\% while prioritizing participant screening procedures. 
These studies define procedural standards that improve the reliability and reproducibility of large-scale webcam-based eye-tracking research.

Despite this progress, it remains unclear which participant- and context-level factors most strongly predict data quality. Our study addresses this gap by modeling how behavioral and technical factors influence data reliability in crowdsourced webcam eye-tracking settings.

\subsection{Applications}
Research on webcam-based eye tracking demonstrates its use for studying attention, cognition, and behavior across three main domains: (1)~\textbf{attention and interface studies}, which examine user engagement and visual saliency; (2)~\textbf{cognitive and linguistic research}, which adapt classic experimental paradigms to online settings; and (3)~\textbf{decision-making and behavioral economics}, which analyze how gaze dynamics influence attention and choice in complex decisions.

\paragraph{Attention and interface studies.}
A major application of webcam-based eye tracking is understanding how people attend to and interact with digital interfaces. \citet{bertrand2023dynamics} examined gaze–cursor coordination during on-screen interaction, while \citet{chen2023examining} analyzed how young adults view e-cigarette marketing materials in realistic web environments. \citet{james2025paradigms} demonstrated that classic attention paradigms can be replicated online, supporting behavioral research despite spatial precision limits. In accessibility research, \citet{edughele2022eye} reviewed gaze-based assistive systems that enable communication and interface control for individuals with motor impairments. \citet{singh2023have} introduced the multimodal \textit{EngageNet} dataset to model user engagement in online learning, and \citet{katsaounidou2025ai} developed the \textit{iMedius} framework to monitor attention to online news and misinformation. \citet{haveriku2025systematic} further showed that eye-movement features enhance linguistic prediction and cross-lingual generalization. In sum, these studies demonstrate how webcam-based eye tracking supports diverse applications in interaction, accessibility, media, and language research.

\paragraph{Cognitive and linguistic research.}
Webcam-based eye tracking is also applied in cognitive and linguistic research. \citet{juantorena2023web} used a web-based prototype for the anti-saccade task, showing that inhibitory control and reaction-time effects can be measured reliably online. \citet{thilderkvist2022current} investigated how programmers read and comprehend source code, finding that gaze patterns and reading linearity differ from natural language and reveal distinct cognitive strategies. \citet{yuksel2025visual} investigated how simulated visual field deficits affect information processing, showing that vision loss increases cognitive load and reduces comprehension through altered gaze behavior. These studies show that webcam-based methods enable remote investigation of executive control, comprehension, and information processing, extending cognitive and linguistic research beyond the laboratory.

\paragraph{Decision-making and behavioral economics.}
Webcam-based eye tracking reveals \textit{how} people make decisions by capturing gaze dynamics as choices unfold.~\citet{yang2021webcam} found that longer and more frequent fixations predict choices and signal decision conflict, supporting the Attentional Drift Diffusion Model (aDDM), which posits that attended information receives greater weight.~\citet{bertrand2023continuous} reported that harder choices elicit longer viewing times and more dwells.~\citet{wong2023experimental} showed that positive versus negative framing of supplier quality shifts attention and purchasing decisions, with gaze mediating this effect. Similarly, \citet{sarvi2025understanding} observed that visually distinct items attract earlier and longer fixations, linking saliency to consumer preference.

While prior application research has used webcam-based eye tracking across behavioral domains, few studies have examined participant-level data quality in these settings. Using data from an AI interview experiment, we model behavioral and technical factors that predict webcam gaze reliability in real-world conditions.

\section{Case Study: Fairness in AI Interviews}

To address \textbf{RQ2}, we analyzed a crowdsourced webcam eye-tracking dataset collected during AI-based job interviews~\citep{lau2026skin}. The original study focused on participants’ trust and fairness perceptions, whereas our goal here is to evaluate the reliability of RealEye’s webcam-based tracking in this socially interactive and unsupervised setting. We model participant data quality by relating RealEye’s quality grade to behavioral and technical factors recorded during the task.

\subsection{Participants}
The final sample comprised 205 valid datasets after excluding incomplete sessions and technical errors from 228 recruited individuals. Participants were adults fluent in English and located primarily in the United States, the United Kingdom, and Germany. All participants self-reported having normal or corrected-to-normal vision. The sample was demographically diverse (mean age $\approx$ 40 years) and representative of the typical heterogeneity of online crowdsourcing studies. No significant demographic differences were observed across participant quality grades. Detailed participant demographics are provided in Appendix Table~\ref{tab:pivoted_demographics}.

\subsection{Dataset and Procedure}

\begin{figure}[!ht]
  \centering
  \begin{minipage}[t]{0.55\textwidth}
    \centering
    \includegraphics[width=\textwidth]{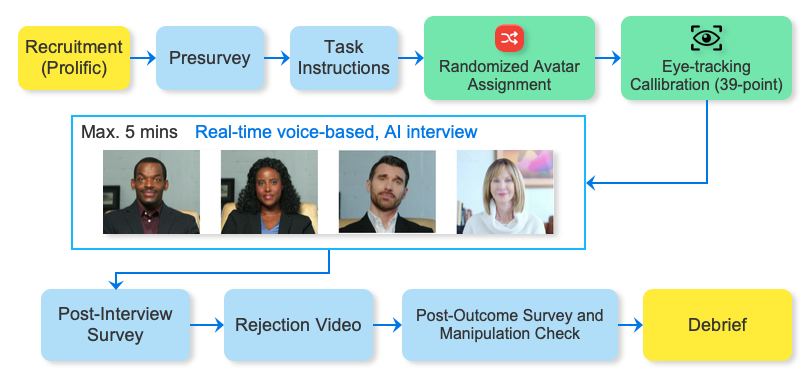}
    \caption{Experimental procedure from recruitment to AI interview.}
    \Description{Flowchart of the experimental procedure. Steps proceed from Recruitment via Prolific~\cite{Prolific}, to Presurvey, Task Instructions, Randomized Avatar Assignment, and 39-point Eye-tracking Calibration. Participants then complete a real-time voice-based AI interview (maximum 5 minutes) with one of four avatar identities. Afterward, they complete a Post-Interview Survey, view a Rejection Video, complete a Post-Outcome Survey and Manipulation Check, and finish with a Debrief.}
    \label{fig:procedure}
  \end{minipage}
  \hfill
  \begin{minipage}[t]{0.4\textwidth}
    \centering
    \includegraphics[width=\textwidth]{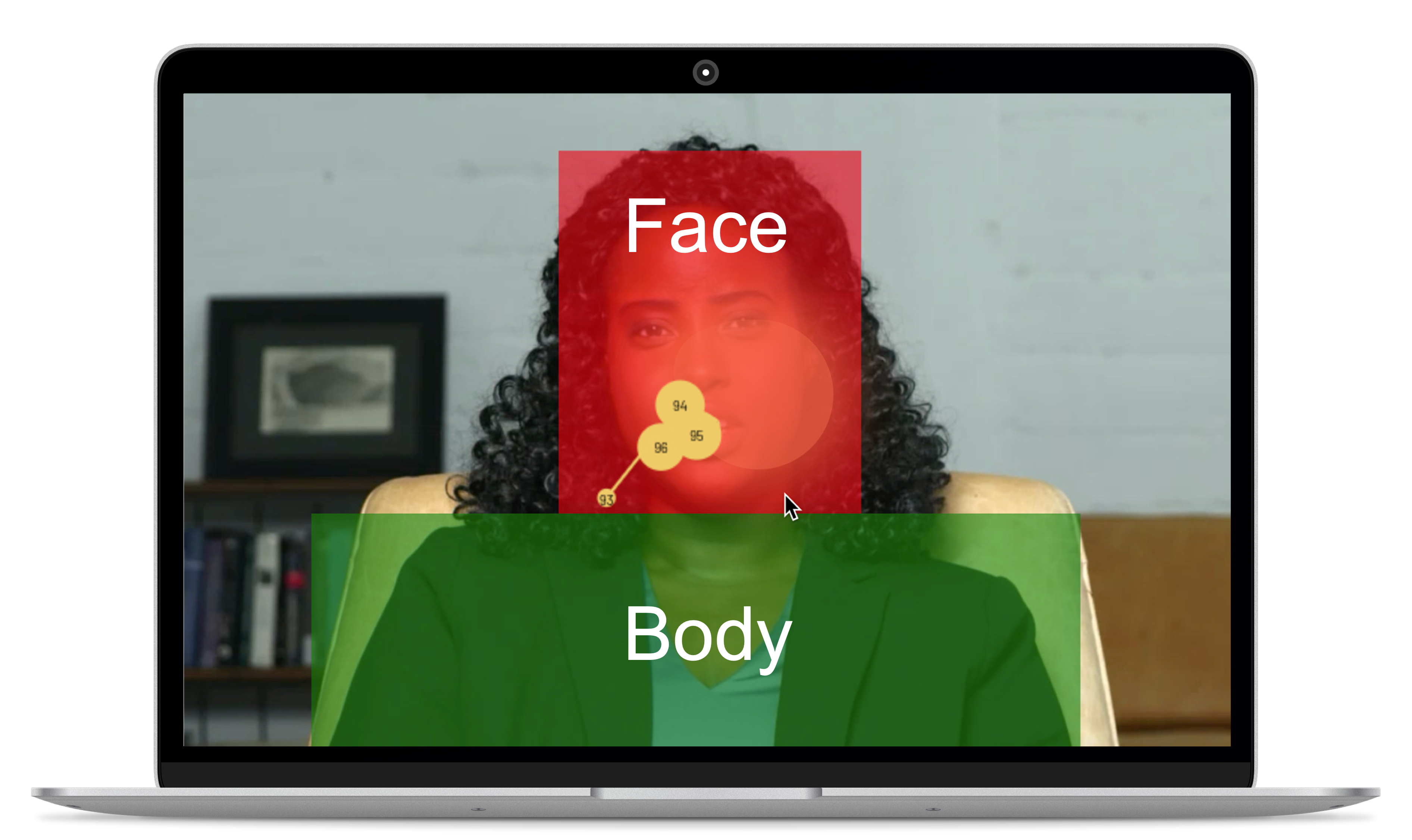}
    \caption{Face and body AOIs defined on the AI interviewer.}
    \Description{Screenshot of the AI interviewer displayed on a laptop screen, with two overlaid areas of interest (AOIs). A red rectangle covers the interviewer's face, labeled Face. A green rectangle covers the interviewer's upper body, labeled Body. A cluster of fixation points is visible near the mouth region.}
    \label{fig:aoi_area}
  \end{minipage}
\end{figure}

We employed a $2\times2$ between-subjects design, manipulating the match or mismatch between participants’ identities and the AI interviewer avatar by race and sex. We chose these avatar categories to reflect a majority–minority contrast motivated by evidence that hiring discrimination is associated with salient visual identity cues such as skin color, which signal perceived cultural distance~\citep{Zschirnt2016}. Participants were recruited via Prolific~\cite{Prolific}, between 2--17 July~2025. Eligibility required English fluency, a functioning webcam and microphone, and stable internet access. Each session lasted approximately 20~minutes, and participants were compensated \pounds4.27 (= \pounds12.80/hour) in accordance with Prolific’s fair-pay policy~\cite{ProlificFairPay}. The overall study workflow, including recruitment, calibration, AI interview, and debriefing, is summarized in Figure~\ref{fig:procedure}. All procedures were approved by the Institutional Review Board (IRB) of the Technical University of Munich, and participants provided informed consent before participating.

\subsection{Webcam Eye-Tracking Setup}

Eye-tracking data were collected using the RealEye platform (version~18.49.0), a browser-based system that estimates gaze position from webcam video at a nominal sampling rate of 10--60~Hz, depending on device performance. The platform performs a 39-point calibration followed by a 3-point validation; participants who failed validation twice were automatically exited and compensated. According to RealEye’s documentation, validation is passed only if the estimated gaze point falls within 150 pixels of each of the three validation targets. RealEye documentation also shows full-screen accuracy is approximately 100--125~pixels, with the highest accuracy in the central region. RealEye does not perform mid-task recalibration using a repeated calibration grid during the study session. Instead, the platform uses a ``Virtual Chinrest'' mechanism to maintain data quality: if a participant moves too far from the calibrated position, they are prompted to return to the correct position before tracking resumes. At present, RealEye does not provide an exportable metric for how many times the virtual chinrest was triggered during a session, which limits our ability to quantify tracking stability at the participant level.

We integrated RealEye’s embedded SDK~\cite{RealEyeSDK} to define two AOIs corresponding to the interviewer’s face and body as shown in Figure~\ref{fig:aoi_area}. We defined only these two regions as AOIs because they are the main socially relevant targets in an interview context. We did not define a background AOI because background content varies across stimuli and is not central to our research questions. To mitigate the known limitations of webcam-based eye trackers in spatial precision and fixation accuracy as mentioned in~\cite{sharafi2020practical, semmelmann2018online, patterson2025methodological}, the AOIs were defined large enough to capture all relevant fixations.

\subsection{Measures and Analysis} 

The outcome variable was the \textbf{Participant Quality Grade} (1 = Very Low to 6 = Perfect), which RealEye computes from four internal signal metrics: sampling rate (Hz),~fixation detection (i.e., a binary indicator of whether the system can compute fixations, requiring 20~Hz sampling rate), eye-tracking data length (also referred to as data integrity), which reflects the extent of gaps in the collected gaze data, and percentage of on-screen gaze time~\cite{RealEyeQuality}. Because RealEye does not provide per-participant gaze estimation error expressed in degrees of visual angle, we rely on the platform's quality grade as our primary outcome measure. Since these signals define the grade itself, we excluded them as predictors to avoid circularity. We instead used other measures provided by RealEye but not used in the grade calculation, along with information we collected separately. 

Predictors were grouped into three categories: (1) \textit{Behavioral factors}: fixation count and test duration (seconds), both exported by RealEye. Since fixation count is derived from event-detection procedures that identify fixations in continuous gaze streams~\cite{kasneci2014applicability}, we interpret it here primarily as a behavioral proxy for participant engagement rather than a direct signal-quality metric. Fixation count comes from RealEye's event-detection pipeline but is not part of the grade formula; removing it worsened model fit by $\Delta$AIC = 120, confirming its value as a predictor of engagement; (2) \textit{Device factors}: browser width (px) and operating system, recorded separately; and (3) \textit{Demographic factors}: participant age, when available in the RealEye metadata.

We report 95\% Wald confidence intervals based on model standard errors. We tested the proportional-odds assumption using a partial proportional-odds (PPO) model that allowed non-parallel effects for \textit{operating system}. The PPO did not improve fit (LR = 10.998, df = 8, $p = .202$; AIC$_{\text{PO}}$ = 543.9 vs. AIC$_{\text{PPO}}$ = 548.9), so we retained the proportional-odds model. Analyses were conducted in Python (version 3.9.6) and RStudio (version 2025.05.1+513).

\subsection{Results: Quality Analysis}

To examine how participant characteristics and session factors relate to data quality, we fitted an OLR model predicting participant quality grade. All predictors showed low multicollinearity (VIFs $< 2$), and model diagnostics indicated adequate fit (AIC = 519.9; McFadden’s pseudo-$R^2 = 0.212$). Table~\ref{tab:orderedmodel} presents the full OLR results. The model identified four statistically significant predictors of participant quality grade: fixation count, test duration, browser width, and operating system. Participant age was not a significant predictor. We describe these results below, organized by behavioral factors, device-related factors, and operating system. To assess robustness, we ran an OLS model predicting mean sampling rate (Hz), reported in Appendix Table~\ref{tab:ols_sampling_rate}. The results confirmed the same direction of effects ($R^2 = 0.64$), indicating that the identified behavioral and device factors predict not only the platform's composite quality grade but also one of its constituent metrics, sampling rate. This convergence suggests that the identified predictors are not artifacts of RealEye’s composite grading scheme.

\begin{table}[!ht]
\small
\centering
\caption{Ordered logistic regression predicting participant quality grade. Positive coefficients indicate a higher likelihood of achieving a higher quality grade.}
\Description{This table reports the results of an ordered logistic regression model predicting participant quality grade. Fixation count has a positive and statistically significant effect, indicating that higher fixation counts are associated with higher quality grades. Test duration has a significant negative effect, suggesting longer durations are associated with lower quality. Browser width and using Mac OS X are associated with higher quality grades. Participant age is not statistically significant. The model includes 205 observations, with a log-likelihood of -250.96, AIC of 519.9, BIC of 549.8, and McFadden's pseudo R-squared of 0.212.}
\label{tab:orderedmodel}
\begin{tabular}{lrrrrrr}
\hline
\textbf{Variable} & \textbf{Coef.} & \textbf{Std. Err.} & \textbf{z} & \textbf{P > |z|}& \textbf{[0.025]} & \textbf{[0.975]} \\
\hline
Fixation Count & 0.0253$^{***}$ & 0.003 & 8.527 & 0.000 & 0.019 & 0.031 \\
Participant Age & -0.0175 & 0.012 & -1.493 & 0.135 & -0.040 & 0.005 \\
Test Duration (s) & -0.0673$^{***}$ & 0.009 & -7.680 & 0.000 & -0.084 & -0.050 \\
Test Browser Width (px) & 0.0013$^{**}$ & 0.000 & 3.054 & 0.002 & 0.000 & 0.002 \\
Operating System (Mac OS X) & 0.7358$^{*}$ & 0.319 & 2.304 & 0.021 & 0.110 & 1.362 \\
\hline
Cutpoints & & & & & & \\
1/2 & -1.5217 & 1.057 & -1.440 & 0.150 & -3.593 & 0.550 \\
2/3 & 0.5509$^{**}$ & 0.186 & 2.959 & 0.003 & 0.186 & 0.916 \\
3/4 & 0.3343$^{*}$ & 0.142 & 2.360 & 0.018 & 0.057 & 0.612 \\
4/5 & 0.5427$^{***}$ & 0.114 & 4.777 & 0.000 & 0.320 & 0.765 \\
\hline
\multicolumn{7}{l}{\textbf{Model fit}} \\
No. of observations & \multicolumn{6}{r}{205} \\
Log-Likelihood & \multicolumn{6}{r}{-250.96} \\
AIC & \multicolumn{6}{r}{519.9} \\
BIC & \multicolumn{6}{r}{549.8} \\
McFadden's pseudo $R^2$ & \multicolumn{6}{r}{0.212} \\
\hline
\multicolumn{7}{l}{Significance: ${}^{*} p < 0.05$, ${}^{**} p < 0.01$, ${}^{***} p < 0.001$} \\
\hline
\end{tabular}
\end{table}

\paragraph{Behavioral factors.}

\begin{figure*}[!ht]
\centering
\begin{subfigure}[t]{0.31\textwidth}
\centering
\includegraphics[width=\textwidth]{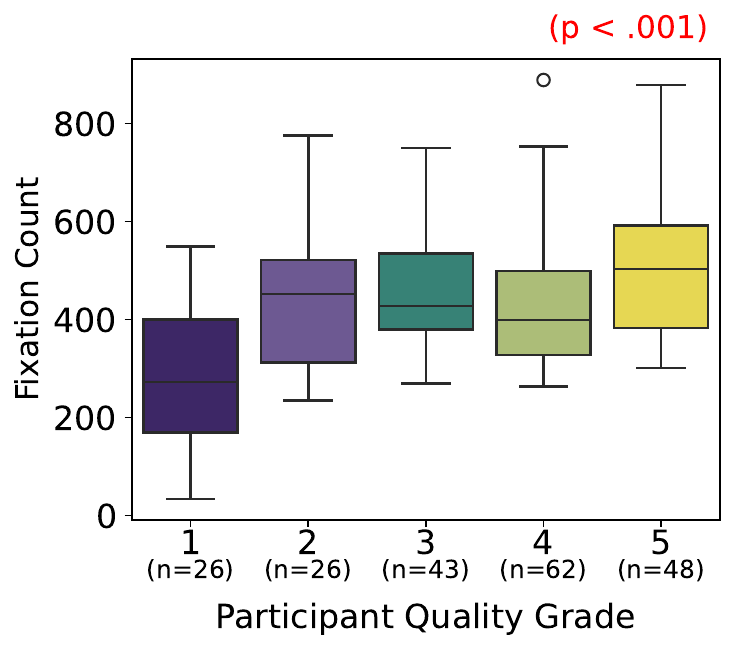}
\caption{Fixation count}
\label{fig:fixation-count}
\end{subfigure}
\hfill
\begin{subfigure}[t]{0.33\textwidth}
\centering
\includegraphics[width=\textwidth]{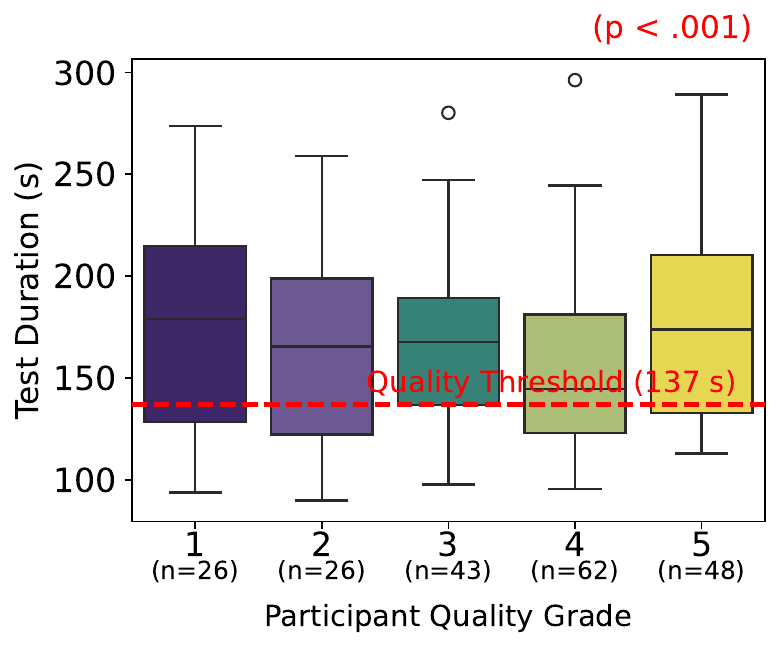}
\caption{Test duration}
\label{fig:test-duration}
\end{subfigure}
\hfill
\begin{subfigure}[t]{0.33\textwidth}
\centering
\includegraphics[width=\textwidth]{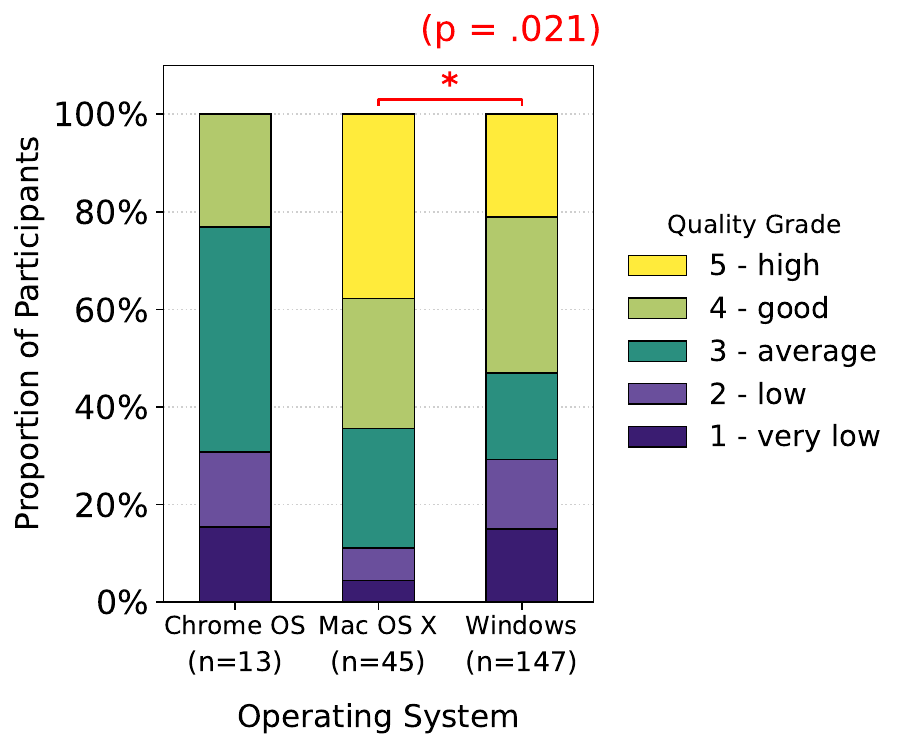}
\caption{Operating system}
\label{fig:operating-system}
\Description{Three panels: (a) Box plot showing fixation count increasing with quality grade (p < .001). (b) Box plot showing test duration decreasing with quality grade (p < .001), with a dashed quality threshold line at 137 seconds. (c) Stacked bar chart showing quality grade distribution by operating system (Chrome OS, Mac OS X, Windows), with Mac OS X having a significantly higher proportion of high-quality grades than Windows (p = .021).}   
\end{subfigure}
\caption{Associations between key predictors and participant quality grade. (a) Fixation count by quality grade. (b) Test duration by quality grade; dashed line indicates the predicted 50\% low-quality threshold ($\approx$137~s). (c) Quality grade distribution by operating system, (*, $p = .021$), asterisks indicate the significant pairwise comparison. Inferential statistics are reported in Table~\ref{tab:orderedmodel}.}
  \label{fig:quality-panels}
\end{figure*}

Fixation count was positively associated with participant data quality, as shown in Table~\ref{tab:orderedmodel} and Figure~\ref{fig:fixation-count}. Each additional fixation increased the odds of belonging to a higher quality grade by about 2.5\% (OR $\approx 1.025$), holding other variables constant. In contrast, longer test durations were associated with lower quality grades, as shown in Table~\ref{tab:orderedmodel} and Figure~\ref{fig:test-duration}. Using the fitted model, we estimated a quality threshold, defined as the test duration at which a participant with average covariates has a 50\% predicted chance of receiving a low-quality grade ($\mathrm{Grade} \le 3$). The estimated threshold is 137 seconds, or about 46\% of the 5-minute eye-tracking period. This value reflects the specifics of our setup and should be viewed as contextual guidance rather than a universal cut-off. The full session, including the interview, pre- and post-interview phases, and the post-outcome questionnaire, lasted about 20 minutes.

\paragraph{Device-related factors.}
Browser width was positively associated with participant quality grade, as shown in Table~\ref{tab:orderedmodel}. Although the effect size is small, this association likely reflects aspects of display geometry and viewing conditions specific to the experimental setup (e.g., differences between desktop monitors and smaller laptop displays).

\paragraph{Operating system.}
The operating system was associated with participant quality grade, as shown in Table~\ref{tab:orderedmodel} and Figure~\ref{fig:operating-system}. Compared to the \textit{Windows} baseline, \textit{Mac OS X} users had higher odds of achieving a better quality grade.

\section{Discussion}

\subsection{Summary of Findings}

Our study addressed two research questions through a scoping review and an empirical quality analysis.~\textbf{RQ1} explored the methodological and validation practices that have influenced webcam-based and crowdsourced eye-tracking research. The scoping review identified three main areas: system development, validation, and application. It showed that, although systems, algorithms, and datasets have advanced rapidly, validation and reporting practices have not kept pace. Few studies report standardized quality metrics or benchmarks, which limits reproducibility across platforms. Most validation work still targets system-level accuracy rather than participant or contextual factors, and procedural standards are inconsistently reported. These gaps highlight the need for predictive modeling approaches, such as this work, that link behavioral and technical variability to measurable data quality. In addressing \textbf{RQ2}, the regression analysis identified behavioral and device-related factors that significantly influenced webcam-based eye-tracking reliability, while demographic variables such as age showed no significant effect.

\paragraph{Behavioral factors.}
Our results show that~\textit{higher fixation count} predicted higher-quality grades, suggesting that attentive and consistent viewing behavior supports more reliable tracking. This finding is consistent with previous reading-based datasets, such as WebQAmGaze~\cite{ribeiro2023webqamgaze}, in which fixation density predicts comprehension accuracy. It is also consistent with our previous task-level validation results, which show that webcam-based eye tracking is most reliable for large AOIs and sustained fixations~\cite{vos2022comparing, prystauka2024online}. Both our results and those of previous studies suggest that fixation count captures both attentional engagement and physical stability, two conditions essential for accurate estimation in uncontrolled environments. In our study, all participants met RealEye’s minimum sampling rate (20 Hz) for fixation detection, suggesting that variation in fixation count reflects differences in viewing behavior rather than detection failures.

Conversely,~\textit{longer test durations} predicted poorer quality.
This pattern aligns with observations that extended testing often results from recalibration attempts or interruptions caused by head movement~\cite{vos2022comparing, patterson2025methodological}. 
While our dataset did not record the number of recalibration events, the RealEye platform includes a virtual chinrest feature that reminds participants to return to the calibrated head position when substantial head movement is detected~\cite{realeye_support}. This mechanism may explain why longer test durations are associated with lower quality grades. In short, test duration acts as an indirect marker of participant fatigue and calibration instability rather than engagement per se.

\paragraph{Device-related factors.} 
Data quality also varied systematically across participants' technical setups.~\textit{Wider browser windows} were associated with higher quality grades, as shown in Table~\ref{tab:orderedmodel}. However, browser width in pixels does not account for physical screen size or pixel density, and may serve as a proxy for overall hardware and display quality (e.g., desktop monitors versus smaller laptops). Since the AOIs in our task were centrally positioned, this effect likely reflects the specific viewing conditions of our interview setup rather than a generalizable relationship between browser width and data quality. Operating-system differences supported prior system-level findings~\cite{vos2022comparing, kaduk2024webcam}. Participants using macOS achieved higher-quality grades, likely due to standardized camera drivers and consistent GPU timing that reduce sampling variability rather than inherent platform superiority~\cite{vos2022comparing, kaduk2024webcam}. Still, interactions between the operating system, browser, and hardware remain the main source of measurement differences in online eye tracking~\cite{prystauka2024online, brandl2024evaluating, uittenhove2022lab}, and in some cases this pattern reverses: mobile users can yield more stable signals due to closer camera distance and steadier lighting~\cite{chen2023examining}. Overall, device effects depend on task context and environment,
reflecting the interplay of the hardware–software ecosystem rather than any single platform.

\paragraph{Demographic factors.} 
Participant demographics such as age showed no significant effect on data quality, consistent with prior webcam- and smartphone-based studies that found no demographic effects on data reliability~\cite{panja2024prediction, ribeiro2023webqamgaze, banki2022comparing}. Most data loss stems from non-compliance, inattention, or head movement rather than participant background~\cite{uittenhove2022lab, banki2022comparing}.

\subsection{Methodological Lessons}

Conducting this study was our first experience combining crowdsourced recruitment with webcam-based eye tracking for a socially interactive AI interview. Inspired by~\citet{burch2024}, we summarize our methodological lessons from two perspectives: that of~\textbf{the researcher} and that of~\textbf{the platform design}.

\paragraph{From the researcher's perspective.}
Our experience showed that conducting webcam-based eye-tracking studies with public participants presents both technical and procedural challenges.
For many first-time users, calibration and sustained tracking required considerable effort. Several participants reported difficulties in the open-ended feedback, such as rapid head movements causing RealEye’s virtual chinrest to reappear (“\textit{The green dots part came up twice and took a long time to complete.}”). Such reports illustrate that while RealEye’s virtual chinrest aims to improve data quality, it can also diminish the participant experience. Future studies should consider incorporating pre-study training or interactive calibration guidance, particularly when working with non-expert participants. For example, researchers may benefit from platforms that offer short animations illustrating appropriate lighting and posture, or simple real-time feedback to support participant self-correction during calibration.

\paragraph{From the platform design perspective.}
Some usability issues arose from interactions between the eye-tracking platform and our experimental interface. In some cases, RealEye’s client-side interface elements interfered with task stimuli, e.g., one participant reported, (“\textit{I had the pop-up for my eyes being in the circle a few times, and it blocked the exit button or covered the interviewer.}”) Such conflicts highlight the need for better integration between eye-tracking platforms and experimental environments that rely on real-time rendering. Furthermore, greater transparency in how RealEye handles data loss would help researchers assess validity and reproducibility.
Access to simple indicators, such as the number of times the virtual chinrest was triggered per participant, would provide valuable insight into tracking stability. We encourage providers to share additional quality metrics to support evidence-based methodological decisions.

\subsection{Recommendations}

Based on our findings and methodological reflections, we propose three recommendations for improving the design and success of crowdsourced webcam eye-tracking studies.

\paragraph{Recommendation 1: Provide clear setup and consent instructions.}
The amount of guidance participants receive directly influences data quality in webcam-based eye-tracking studies. 
As noted in previous work~\cite{banki2022comparing, semmelmann2018online, patterson2025methodological}, researchers should provide clear, step-by-step instructions on device setup, lighting, and head stability, along with transparent explanations of data use and privacy protection. In our study, participants were informed that “\textit{no video or audio recordings are stored, only gaze data are processed.}” and completed a short checklist to verify webcam functionality, internet stability, and calibration readiness. Clear preparation reduces tracking loss, ensures fair participation, and fosters trust in remote data collection. To support reproducibility, a concise reporting checklist for webcam-based eye-tracking studies is provided in the Appendix Table~\ref{tab:checklist-webcam-et}.

\paragraph{Recommendation 2: Evaluate platform architecture and integration early.} Webcam eye-tracking platforms differ in how they handle data collection and calibration. Some platforms rely on screen recording or browser-based gaze estimation, while others embed experiments directly into their interfaces. These differences affect experimental control and data access. In our study, RealEye’s embedded design limited control over calibration and interface rendering. 
Researchers should assess these constraints early to ensure the chosen platform aligns with their experimental goals.

\paragraph{Recommendation 3: Operationalize data quality for screening and exclusion.}
Researchers can operationalize data quality using behavioral indicators such as fixation count or test duration as post-hoc screening variables. For example, unusually long sessions or very low fixation counts may signal tracking instability or participant fatigue and can be flagged for sensitivity analyses or excluded based on pre-registered criteria. These predictors, identified in our analysis, provide practical guidance for defining quality thresholds in crowdsourced webcam-based eye-tracking studies.

\subsection{Limitations and Future Work}

\paragraph{Missing spatial accuracy metrics.}

RealEye does not provide per-participant gaze estimation error in degrees of visual angle, nor does it export raw validation offsets. Based on the platform's reported accuracy of approximately 106 pixels in the initial validation task~\cite{RealEyeWhitePaper} and assumptions about typical viewing distances (50–60 cm) and screen pixel densities (0.17–0.27 mm/px), we estimated spatial accuracy at $1.7^\circ$ to $3.3^\circ$. Because neither viewing distance nor physical screen size were recorded, this estimate should be interpreted as an approximation rather than a participant-level measure. We encourage platform providers to report, for each participant, the mean gaze offset observed during validation, expressed in degrees of visual angle, to support independent quality assessment.

\paragraph{Data quality was not linked to task performance.}
We found no significant associations between quality grade and interview duration or transcript word count (Spearman, $N=205$, all $p>.40$; Kruskal–Wallis, all $p>.10$), though these proxies may
not capture meaningful performance variation given the absence of well-defined accuracy outcomes. Prior work has shown that spatial inaccuracies can distort fixation-based measures such as dwell
time~\cite{Holmqvist2012, brandl2024evaluating}, and that data quality moderates correlations between reading behavior and model predictions~\cite{morger2022}. Future studies could correlate participants' attention check pass rates on the recruitment platform with their eye-tracking quality grades to assess whether recruitment-level screening predicts tracking reliability in crowdsourced settings.

\paragraph{Platform and task generalizability.}
The OLR results are specific to RealEye’s algorithms and quality scoring system and may not transfer directly to other webcam-based platforms. Likewise, our AI interview task involved a limited set of participants' demographic and avatar identities. While these factors constrain generalization, the modeling approach provides a replicable template for identifying predictors of data quality across systems and study designs. Future work could test whether these predictors generalize across platforms and tasks, for example, by replicating the analysis with other commercial webcam eye-tracking services or task types (e.g., reading or visual search).

\paragraph{Sample characteristics.}
In our experiment, the distribution of quality grades was uneven, with few participants at the extremes. For the analysis, we merged Grades~5 and~6 due to the small highest-grade sample ($n=3$). Model diagnostics confirmed adequate stability, but future research could ensure more balanced sampling across quality levels through stratified recruitment. Additionally, our study did not oversample to compensate for data loss due to calibration failures, a common issue in crowdsourced webcam eye-tracking. Following recent recommendations~\cite{patterson2025methodological}, future studies could increase sample sizes by 20–40\% beyond a priori power estimates.

\section{Conclusion}

This work combined a scoping review and a case study to evaluate factors influencing participant data quality in crowdsourced, webcam-based eye tracking. The review revealed fragmented reporting practices and limited consideration of behavioral and technical factors that affect data reliability. In our case study, analysis of data collected through the RealEye platform showed that fixation count, test duration, and operating system were consistent predictors of participant-level quality. These findings provide an empirical basis for assessing and improving remote gaze data. We encourage transparent reporting standards, reproducible analysis pipelines, and platform-independent quality models to enhance the reliability and comparability of webcam-based eye tracking in future research.

\section{Societal Impact Statement}
Webcam-based eye tracking offers scalable access to diverse participants but also raises concerns about privacy and data sovereignty. In this study, only gaze coordinates were recorded (no video), and all procedures followed IRB-approved consent and fair-compensation standards. We emphasize transparent data handling and advocate open reporting to promote privacy-conscious and ethical practices in future crowdsourced webcam-based eye-tracking research.

\section{Open Science}
To support reproducibility, we provide the complete analysis pipeline and scripts used for data processing at~\url{https://gitlab.lrz.de/hctl/crowdsourced-webcam-eyetracking-analysis}.

\bibliographystyle{ACM-Reference-Format}
\bibliography{main}

\appendix
\clearpage

\section{Participant Demographics}

\begin{table}[!ht]
\centering
\footnotesize
\caption{Participant demographics and individual differences by quality grade (N = 205). Values represent means $\pm$ standard deviations, or counts as indicated.}
\Description{This table summarizes participant demographics across five quality grades (1 to 5) for a total of 205 participants. Each row corresponds to a quality grade and reports gender distribution, mean age, ethnicity, vision status, employment rate, prior interview experience, and self-reported nervousness. Across grades, gender distribution varies, with both men and women represented in each group. Mean age ranges from approximately 37 to 42 years. Ethnicity distribution shows variation in counts of White and Black participants across grades. Most participants report normal vision or use glasses, with few using contact lenses. Employment rates are high across all groups, generally above 80 percent. A majority of participants report moderate or extensive prior interview experience. Self-reported nervousness levels are similar across groups, with mean values around 4 on a 1 to 10 scale.}
\label{tab:pivoted_demographics}
\begin{tabular}{cccccccc}
\toprule
\makecell{\textbf{Quality Grade}\\(n)} &
\makecell{\textbf{Gender}\\(M/W)} &
\makecell{\textbf{Age}\\(years)} &
\makecell{\textbf{Ethnicity}\\(W/B)} &
\makecell{\textbf{Vision}\\(N/G/C)} &
\makecell{\textbf{Employment}\\(\%)} &
\makecell{\textbf{Interview}\\ (Exp. \%)} &
\makecell{\textbf{Nervousness}\\(1--10)} \\
\midrule
1 (n=26) & 10/16 & 42.2 $\pm$ 11.1 & 18/8  & 17/9/0  & 85 & 58 & 4.54 $\pm$ 2.76 \\
2 (n=26) & 17/9  & 42.2 $\pm$ 11.5 & 12/14 & 16/9/1  & 92 & 81 & 4.12 $\pm$ 2.82 \\
3 (n=43) & 19/24 & 38.3 $\pm$ 10.5 & 14/29 & 29/11/3 & 91 & 79 & 4.00 $\pm$ 2.47 \\
4 (n=62) & 28/34 & 40.9 $\pm$ 11.9 & 36/26 & 45/14/3 & 90 & 73 & 4.02 $\pm$ 2.55 \\
5 (n=48) & 28/20 & 37.1 $\pm$ 11.9 & 27/21 & 29/17/2 & 81 & 71 & 4.21 $\pm$ 2.77 \\ 
\bottomrule
\end{tabular}

\vspace{0.5em}
\raggedright
{\footnotesize
\textit{Notes.} Gender (M/W) = Men/Women; Ethnicity (W/B) = White/Black; Vision (N/G/C) = Normal/Glasses/Contact lenses;
Employment values indicate the percentage of participants employed full- or part-time;
Interview experience values indicate the percentage reporting moderate or extensive prior interview experience.
}
\end{table}

\section{Robustness Check}

\begin{table}[!ht]
\small
\centering
\caption{OLS regression predicting mean sampling rate (Hz). The model uses the same predictors as the main OLR (Table~\ref{tab:orderedmodel}) to assess robustness. Positive coefficients indicate higher sampling rates.}
\Description{This table reports the results of an ordinary least squares regression predicting mean sampling rate in Hertz using the same predictors as the main ordered logistic regression model. Fixation count has a positive and statistically significant association with sampling rate, indicating that higher fixation counts correspond to higher sampling rates. Test duration has a significant negative effect, suggesting that longer tests are associated with lower sampling rates. Browser width has a positive and significant effect. Participant age and operating system (Mac OS X) are not statistically significant predictors. The model includes 205 observations and explains a substantial proportion of variance, with an R-squared of 0.642 and an adjusted R-squared of 0.633.}
\label{tab:ols_sampling_rate}
\begin{tabular}{lrrrrrr}
\hline
\textbf{Variable} & \textbf{Coef.} & \textbf{Std. Err.} & \textbf{t} & \textbf{P > |t|} & \textbf{[0.025]} & \textbf{[0.975]} \\
\hline
Fixation Count & 0.0513$^{***}$ & 0.003 & 18.506 & 0.000 & 0.046 & 0.057 \\
Participant Age & 0.0100 & 0.023 & 0.427 & 0.670 & -0.036 & 0.056 \\
Test Duration (s) & -0.1354$^{***}$ & 0.009 & -15.032 & 0.000 & -0.153 & -0.118 \\
Test Browser Width (px) & 0.0031$^{***}$ & 0.001 & 3.556 & 0.000 & 0.001 & 0.005 \\
Operating System (Mac OS X) & 0.8352 & 0.655 & 1.274 & 0.204 & -0.457 & 2.128 \\
\hline
\multicolumn{7}{l}{\textbf{Model fit}} \\
No. of observations & \multicolumn{6}{r}{205} \\
Log-Likelihood & \multicolumn{6}{r}{-561.91} \\
AIC & \multicolumn{6}{r}{1136} \\
BIC & \multicolumn{6}{r}{1156} \\
$R^2$ & \multicolumn{6}{r}{0.642} \\
Adj. $R^2$ & \multicolumn{6}{r}{0.633} \\
\hline
\multicolumn{7}{l}{Significance: ${}^{*} p < 0.05$, ${}^{**} p < 0.01$, ${}^{***} p < 0.001$} \\
\hline
\end{tabular}
\end{table}

\section{Reporting Checklist}

\noindent The following checklist summarizes essential reporting items for planning, conducting, and documenting crowdsourced webcam-based eye-tracking experiments.

\begin{table}[!ht]
\centering
\footnotesize
\caption{Checklist for conducting crowdsourced webcam-based eye-tracking studies.}
\Description{This table presents a checklist for conducting crowdsourced webcam-based eye-tracking studies, organized into seven categories: participant setup, platform configuration, data quality metrics, behavioral measures, device and context, demographics and ethics, and reproducibility. Each category contains a list of recommended reporting items. Participant setup includes instructions on hardware compatibility, calibration readiness, and participant preparation. Platform configuration covers details about the eye-tracking system, calibration procedures, and sampling rates. Data quality metrics include definitions of quality indicators, thresholds, and handling of missing data. Behavioral measures address fixation detection methods, experiment duration, and areas of interest. Device and context include system and display characteristics. Demographics and ethics cover participant information, consent, and approvals. Reproducibility includes sharing of code, data, and preprocessing procedures.}
\label{tab:checklist-webcam-et}
\begin{tabular}{p{0.28\linewidth} p{0.62\linewidth}}
\toprule
\small
\textbf{Category} & \textbf{Checklist Item} \\
\midrule

\textbf{Participant Setup} &
\begin{minipage}[t]{\linewidth}\vspace{2pt}
\begin{itemize}[leftmargin=*, itemsep=2pt, topsep=0pt]
  \item Confirm webcam and browser compatibility before participation.
  \item Provide clear setup instructions (lighting, posture, viewing distance).
  \item Inform participants about data use and privacy.
  \item Verify calibration readiness and camera permissions.
  \item Screen for visual or hardware limitations that affect tracking.
\end{itemize}\vspace{2pt}
\end{minipage} \\
\midrule

\textbf{Platform Configuration} &
\begin{minipage}[t]{\linewidth}\vspace{2pt}
\begin{itemize}[leftmargin=*, itemsep=2pt, topsep=0pt]
  \item Specify the platform provider, version, and gaze estimation mode.
  \item Describe calibration and validation procedures with pass/fail criteria.
  \item Report nominal and measured sampling rates.
  \item Indicate how recalibration or data loss is handled automatically.
\end{itemize}\vspace{2pt}
\end{minipage} \\
\midrule

\textbf{Data Quality Metrics} &
\begin{minipage}[t]{\linewidth}\vspace{2pt}
\begin{itemize}[leftmargin=*, itemsep=2pt, topsep=0pt]
\item Define the quality indicator used (platform-provided or user-defined).
\item Report inclusion and exclusion thresholds and handling of missing data.
\item Report key signal metrics that contribute to data quality (e.g., sampling rate, valid-sample, gaze-on-screen).

\end{itemize}\vspace{2pt}
\end{minipage} \\
\midrule

\textbf{Behavioral Measures} &
\begin{minipage}[t]{\linewidth}\vspace{2pt}
\begin{itemize}[leftmargin=*, itemsep=2pt, topsep=0pt]
\item Describe event-detection algorithms and parameters~\cite{tafaj2013online}.
\item Report session duration, gaze recording length, and engagement proxies.
\item Provide AOI definitions and any normalization method used.
\end{itemize}\vspace{2pt}
\end{minipage} \\
\midrule

\textbf{Device and Context} &
\begin{minipage}[t]{\linewidth}\vspace{2pt}
\begin{itemize}[leftmargin=*, itemsep=2pt, topsep=0pt]
  \item Record operating system and browser (name and version).
  \item Record viewport or browser window dimensions (px).
  \item Note display mode (fullscreen vs.\ windowed) and any zoom/scaling.
\end{itemize}\vspace{2pt}
\end{minipage} \\
\midrule

\textbf{Demographics and Ethics} &
\begin{minipage}[t]{\linewidth}\vspace{2pt}
\begin{itemize}[leftmargin=*, itemsep=2pt, topsep=0pt]
  \item Report demographic variables collected and screening criteria.
  \item Describe informed consent, compensation, and data-retention policy.
  \item State IRB or ethics approval reference.
\end{itemize}\vspace{2pt}
\end{minipage} \\
\midrule

\textbf{Reproducibility} &
\begin{minipage}[t]{\linewidth}\vspace{2pt}
\begin{itemize}[leftmargin=*, itemsep=2pt, topsep=0pt]
  \item Share code, data dictionary, and analysis scripts if permitted.
  \item Note preprocessing or exclusion scripts used for quality control.
\end{itemize}\vspace{2pt}
\end{minipage} \\
\bottomrule
\end{tabular}
\end{table}

\end{document}